# AMPLITUDE ZEROS AND RAPIDITY CORRELATIONS IN $W\gamma$ AND $WZ$ PRODUCTION IN HADRONIC COLLISIONS[*]

U. BAUR

*Physics Department, Florida State University, Tallahassee, FL 32306, USA*

## ABSTRACT

A comparative study of amplitude zeros in $W\gamma$ and $WZ$ production in hadronic collisions is presented. The Standard Model amplitude for $q_1\bar{q}_2 \to W^\pm Z$ at the Born-level is shown to exhibit an approximate zero located at $\cos\theta = (g_-^{q_1} + g_-^{q_2})/(g_-^{q_1} - g_-^{q_2})$ at high energies, where the $g_-^{q_i}$ ($i = 1, 2$) are the left-handed couplings of the $Z$-boson to quarks and $\theta$ is the center of mass scattering angle of the $W$-boson. This approximate zero is similar to the well-known radiation zero in $W\gamma$ production. Prospects to observe the amplitude zeros using rapidity correlations between the final state particles are explored.

## 1. Introduction

Although the electroweak Standard Model (SM) based on an $SU_L(2) \otimes U_Y(1)$ gauge theory has been very successful in describing contemporary high energy physics experiments, the three vector-boson couplings predicted by this non-Abelian gauge theory remain largely untested experimentally. Careful studies of these couplings, for example in di-boson production in $e^+e^-$ or hadronic collisions, may allow us to test the non-Abelian gauge structure of the SM.[1] Constraints on anomalous $WW\gamma$, $WWZ$, $ZZ\gamma$ and $Z\gamma\gamma$ couplings have been reported at this conference by the CDF,[2] DØ[3] and L3 collaboration.[4]

## 2. Amplitude Zeros

The reactions $p\overset{(-)}{p} \to W^\pm \gamma$ and $p\overset{(-)}{p} \to W^\pm Z$ are of special interest due to the presence of amplitude zeros. It is well known that all SM helicity amplitudes of the parton-level subprocess $q_1\bar{q}_2 \to W^\pm \gamma$ vanish for $\cos\theta = (Q_1 + Q_2)/(Q_1 - Q_2)$,[5] where $\theta$ is the scattering angle of the $W$-boson with respect to the quark ($q_1$) direction, and $Q_i$ ($i = 1, 2$) are the quark charges in units of the proton electric charge $e$. This zero is a consequence of the factorizability[6] of the amplitudes in gauge theories into one factor which contains the gauge coupling dependence and another which contains spin information. Although the factorization holds for any four-particle Born-level amplitude

---

[*]Talk given at the DPF'94 Conference, Albuquerque, New Mexico, August 2 – 6, 1994, to appear in the proceedings

in which one or more of the four particles is a gauge-field quantum, the amplitudes for most processes may not necessarily develop a kinematical zero in the physical region. The amplitude zero in the $W^\pm\gamma$ process has been further shown to correspond to the absence of dipole radiation by colliding particles with the same charge-to-mass ratio,[7] a realization of classical radiation interference.

Recently, it was found[8] that the SM amplitude of the process $q_1\bar{q}_2 \to W^\pm Z$ also exhibits an approximate zero at high energies. The $(\pm, \mp)$ amplitudes $\mathcal{M}(\pm, \mp)$ vanish for

$$\frac{g_-^{q_1}}{\hat{u}} + \frac{g_-^{q_2}}{\hat{t}} = 0, \tag{1}$$

where $g_-^{q_i}$ is the coupling of the $Z$ boson to left-handed quarks, and $\hat{s}$, $\hat{u}$ and $\hat{t}$ are Mandelstam variables in the parton center of mass frame. For $\hat{s} \gg M_Z^2$, the zero in the $(\pm, \mp)$ amplitudes is located at $\cos\theta_0 = (g_-^{q_1} + g_-^{q_2})/(g_-^{q_1} - g_-^{q_2})$, or

$$\cos\theta_0 \simeq \begin{cases} +\frac{1}{3}\tan^2\theta_w \simeq +0.1 & \text{for } d\bar{u} \to W^-Z, \\ -\frac{1}{3}\tan^2\theta_w \simeq -0.1 & \text{for } u\bar{d} \to W^+Z. \end{cases}$$

The existence of the zero in $\mathcal{M}(\pm, \mp)$ at $\cos\theta_0$ is a direct consequence of the contributing Feynman diagrams and the left-handed coupling of the $W$-boson to fermions.

At high energies, strong cancellations occur, and, besides $\mathcal{M}(\pm, \mp)$, only the $(0,0)$ amplitude remains non-zero. The combined effect of the zero in $\mathcal{M}(\pm, \mp)$ and the gauge cancellations at high energies in the remaining helicity amplitudes results in an approximate zero for the $q_1\bar{q}_2 \to W^\pm Z$ differential cross section at $\cos\theta \approx \cos\theta_0$. This is illustrated in Fig. 1a where we show the differential cross sections for $d\bar{u} \to W^-Z$ for $(\lambda_w, \lambda_z) = (\pm, \mp)$ and $(0,0)$, as well as the unpolarized cross section, which is obtained by summing over all $W$ and $Z$ boson helicity combinations (solid line). Although the matrix elements are calculated at $\sqrt{\hat{s}} = 2$ TeV, the results differ little from those obtained in the high energy limit. The total differential cross section displays a pronounced minimum at the location of the zero in $\mathcal{M}(\pm, \mp)$.

Figure 1b illustrates the energy dependence of the differential cross section. At $\sqrt{\hat{s}} = 0.2$ TeV, *i.e.* close to the threshold, contributions from the $(\pm, \pm)$, $(\pm, 0)$, and $(0, \pm)$ amplitudes are important. Above threshold, these contributions rapidly diminish, as exemplified by the curves for $\sqrt{\hat{s}} = 0.5$ TeV (dashed lines) and $\sqrt{\hat{s}} = 2$ TeV (solid lines). Note that the location of the minimum varies only slightly with energy.

## 3. Rapidity Correlations

The radiation zero in $q_1\bar{q}_2 \to W\gamma$ and the approximate amplitude zero in $q_1\bar{q}_2 \to WZ$ are not easy to observe in the $\cos\theta$ distribution in $pp$ or $p\bar{p}$ collider experiments. Structure function effects transform the zero in the $W\gamma$ case into a dip in the $\cos\theta$ distribution. The approximate zero in $WZ$ production is only slightly affected by structure function effects. Higher order QCD corrections and finite $W$ width effects tend to fill in the dip. In $W\gamma$ production photon radiation from the final state lepton line also diminishes the significance of the dip. Finally, finite detector resolution effects and ambiguities in reconstructing the parton center of mass frame represent an additional major complication in the extraction of the $\cos\theta$ distribution, and further dilute the signal of the amplitude zeros. The ambiguities are associated with the nonobservation of the neutrino arising from $W$ decay. Identifying the missing transverse momentum with the transverse momentum of the neutrino of a given $W\gamma$ or $WZ$ event, the unobservable longitudinal neutrino momentum, $p_L(\nu)$, and thus the parton center of mass frame, can be reconstructed by imposing the constraint that the neutrino and charged lepton four momenta combine to form the $W$ rest mass.[9] The resulting quadratic equation, in general, has two solutions. In the approximation of a zero $W$ decay width, one of the two solutions coincides with the true $p_L(\nu)$. On an event to event basis, however, it is impossible to tell which of the two solutions is the correct one. This ambiguity considerably smears out the dip caused by the amplitude zeros.

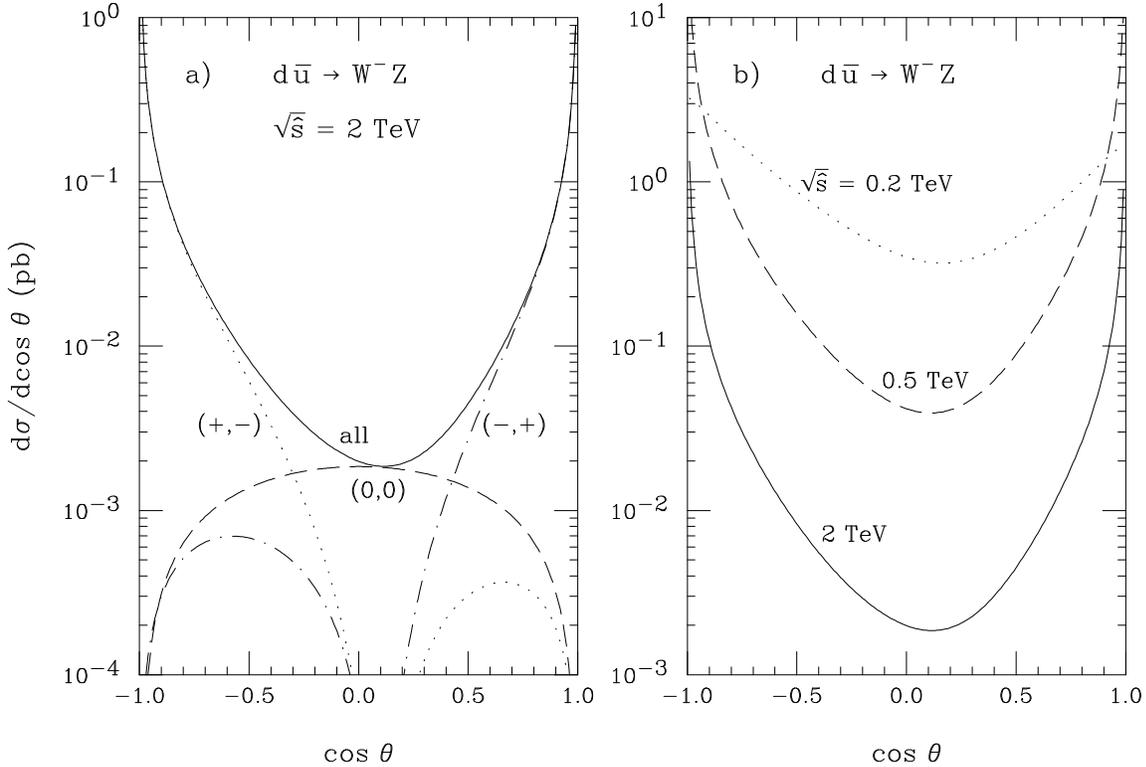

Figure 1: Differential cross section $d\sigma/d\cos\theta$ versus the $W^-$ scattering angle $\theta$ in the center of mass frame for the Born-level process $d\bar{u} \to W^- Z$. a) The contributions of the individual helicity amplitudes, together with the total differential cross section (solid line), are shown for $\sqrt{\hat{s}} = 2$ TeV. b) The differential cross section for three different parton center of mass energies.

Instead of trying to reconstruct the parton center of mass frame and measure the $\cos\theta$ or the equivalent rapidity distribution in the center of mass frame, one can study rapidity correlations between the observable final state particles in the laboratory frame.[10] Knowledge of the neutrino longitudinal momentum is not required in determining these correlations. Event mis-reconstruction problems originating from the two

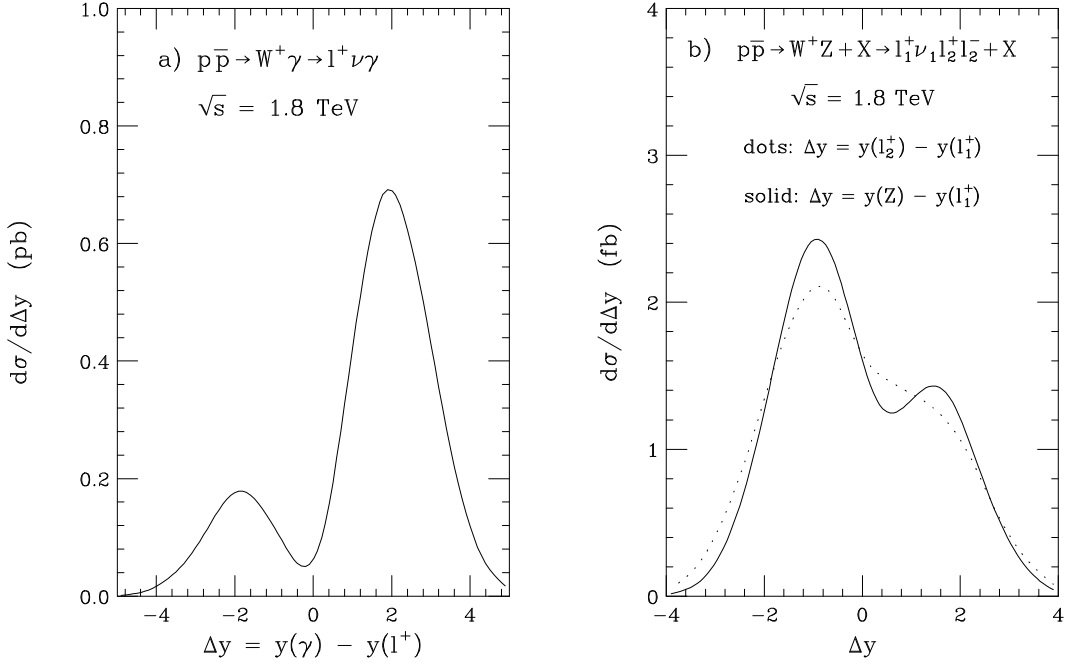

Figure 2: Rapidity difference distributions in the SM at the Tevatron. a) The photon lepton rapidity difference spectrum in $p\bar{p} \to \ell^+ \not{p}_T \gamma$. b) The $y(Z) - y(\ell_1^+)$ and $y(\ell_2^+) - y(\ell_1^+)$ distributions in $p\bar{p} \to W^+ Z$.

possible solutions for $p_L(\nu)$ are thus automatically avoided.

In $2 \to 2$ reactions differences of rapidities are invariant under boosts. One therefore expects that the (lab. frame) rapidity difference distributions $d\sigma/d\Delta y(V, W)$, $V = \gamma, Z$, where $\Delta y(V, W) = y(V) - y(W)$, exhibit a dip signaling the SM amplitude zeros.[10] In $W^{\pm}\gamma$ production, the dominant $W$ helicity is $\lambda_W = \pm 1$,[11] implying that the charged lepton, $\ell = e, \mu$, from $W \to \ell\nu$ tends to be emitted in the direction of the parent $W$, and thus reflects most of its kinematic properties. As a result, the dip signaling the SM radiation zero should manifest itself in the $\Delta y(\gamma, \ell) = y(\gamma) - y(\ell)$ distribution.

The $\Delta y(\gamma, \ell)$ differential cross section for $p\bar{p} \to \ell^+ \not{p}_T \gamma$ at the Tevatron is shown in Fig. 2a. To simulate detector response, transverse momentum cuts of $p_T(\gamma) > 5$ GeV, $p_T(\ell) > 20$ GeV and $\not{p}_T > 20$ GeV, rapidity cuts of $|y(\gamma)| < 3$ and $|y(\ell)| < 3.5$, a cluster transverse mass cut of $M_T(\ell\gamma; \not{p}_T) > 90$ GeV and a lepton photon separation cut of $\Delta R(\gamma, \ell) > 0.7$ have been imposed. The SM radiation zero is seen to lead to a strong dip in the $\Delta y(\gamma, \ell)$ distribution at $\Delta y(\gamma, \ell) \approx -0.3$.

In contrast to the situation in $W\gamma$ production, none of the $W$ helicities dominates in $WZ$ production.[11] The charged lepton originating from the $W$ decay, $W \to \ell_1\nu_1$, thus only partly reflects the kinematical properties of the parent $W$ boson. As a result, a significant part of the correlation present in the $y(Z) - y(W)$ spectrum is lost, and only a slight dip survives in the $y(Z) - y(\ell_1)$ distribution, which is shown for the $W^+Z$ case in Fig. 2b. Experimentally, the $Z$ boson rapidity, $y(Z)$, can readily be reconstructed from the four momenta of the lepton pair $\ell_2^+\ell_2^-$ originating from the $Z$ decay. The cuts used in Fig. 2b are the same as those in Fig. 2a except for the lepton rapidity cut which has been replaced by $|y(\ell_{1,2})| < 2.5$.

Instead of the $y(Z) - y(\ell_1)$ distribution, one could also think of studying the rapidity correlations between the charged leptons originating from the $Z \to \ell_2^+\ell_2^-$ and $W \to \ell_1\nu_1$ decays. The dotted line in Fig. 2b shows the $y(\ell_2^+) - y(\ell_1^+)$ distribution for $W^+Z$ production at the Tevatron. The $y(\ell_2^-) - y(\ell_1^+)$ spectrum almost coincides with the $y(\ell_2^+) - y(\ell_1^+)$ distribution. Since also none of the $Z$ boson helicities dominates[11] in $q_1\bar{q}_2 \to WZ$, the rapidities of the leptons from $W$ and $Z$ decays are almost completely uncorrelated, and essentially no trace of the dip signaling the approximate amplitude zero is left in the $y(\ell_2^+) - y(\ell_1^+)$ distribution.

## 4. Conclusions

$W\gamma$ and $WZ$ production in hadronic collisions are of special interest due to the presence of amplitude zeros in the physical region. These amplitude zeros are best observed in rapidity correlations between the final state particles. In $W\gamma$ production at the Tevatron, the radiation zero leads to a pronounced dip in the $y(\gamma) - y(\ell)$ distribution. In $WZ$ production, the approximate amplitude zero is signalled by a dip in the analogous $y(Z) - y(\ell_1)$ distribution where $\ell_1$ is the charged lepton originating from the $W$ decay. However, the dip in the $y(Z) - y(\ell_1)$ distribution is much less pronounced than that in the $y(\gamma) - y(\ell)$ spectrum in $W\gamma$ production, and thus more difficult to observe experimentally.

## 5. Acknowledgement

This research was supported by the U.S. Department of Energy under Contract No. DE-FG05-87ER40319.